\documentclass{article}
\usepackage{spconf,amsmath,graphicx, tikz}
\usepackage{pgfplots, pgfplotstable}
\usepackage{mathrsfs}
\usepackage{subfigure}
\usepackage{multirow}
\usepackage{amssymb}
\usepackage{booktabs}

\title{TGAVC: Improving Autoencoder Voice Conversion with \\ Text-Guided and Adversarial Training}
\name{Huaizhen Tang$^{1,2}$, Xulong Zhang$^1$, Jianzong Wang$^{1\ast}$\thanks{$^{\ast}$Corresponding author: Jianzong Wang, jzwang@188.com. This paper is supported by National Key Research and Development Program of China under grant No. 2018YFB0204403, No. 2017YFB1401202 and No. 2018YFB1003500.}, Ning Cheng$^1$, Zhen Zeng$^1$, Edward 
Xiao$^3$, Jing Xiao$^1$}\address{$^1$Ping An Technology (Shenzhen) Co., Ltd.\\$^2$ University of Science and Technology of China. \\$^3$ Aquinas International Academy, CA, USA  }

\begin{document}

\maketitle
\begin{abstract}
    Non-parallel many-to-many voice conversion remains an interesting but challenging speech processing task. Recently, AutoVC, a conditional autoencoder based method, achieved excellent conversion results by disentangling the speaker identity and the speech content using information-constraining bottlenecks. However, due to the pure autoencoder training method, it is difficult to evaluate the separation effect of content and speaker identity. In this paper, a novel voice conversion framework, named $\boldsymbol T$ext $\boldsymbol G$uided $\boldsymbol A$utoVC(TGAVC), is proposed to more effectively separate content and timbre from speech, where an expected content embedding produced based on the text transcriptions is designed to guide the extraction of voice content. In addition, the adversarial training is applied to eliminate the speaker identity information in the estimated content embedding extracted from speech. Under the guidance of the expected content embedding and the adversarial training, the content encoder is trained to extract speaker-independent content embedding from speech. Experiments on AIShell-3 dataset show that the proposed model outperforms AutoVC in terms of naturalness and similarity of converted speech.
\end{abstract}


\section{Introduction}

Voice conversion (VC) aims to modify the voice characteristic and speaking style of a source speech while preserving the linguistic information \cite{VoiceConversion1989,stylianou1998continuous}. According to the characteristic of training data, these methods can be roughly categorized into two class, i.e. parallel VC and non-parallel VC \cite{VC-Overview}. More researchers focus on the solutions of non-parallel VC due to the lack of paired source-target speech datasets which is required for parallel VC. 

In recent years, due to the advance of deep learning, various voice conversion solutions are proposed. Since the task of voice conversion is very similar to that of image style transfer in computer vision, many generative adversarial networks (GAN) that have been successfully applied in the field of image conversion also were been adapted to achieve voice conversion, such as CycleGAN-VC \cite{CycleGAN-VC}, StarGAN-VC \cite{StarGAN-VC}. These GAN-based models jointly train a generator network with a discriminator, where an adversarial loss derived from the discriminator is used to encourage the generator to build outputs that are indistinguishable from real speech. Due to the design of the cycle consistency training, GAN-based VC models can be trained with non-parallel speech datasets. In CycleGAN-VC \cite{CycleGAN-VC}, an identity mapping loss is designed to force the generator to find the mapping that preserves the linguistic information. CycleGAN-VC2 \cite{CycleGAN-VC2} proposes two-step adversarial losses and improved the network structure for the generator and discriminator. 
In StarGAN-VC \cite{StarGAN-VC}, many-to-many conversion is achieved through using a speaker identity vector as an additional input of the generator to control the generated speech identity. And, GAZEV \cite{GAZEVs} improved the VC performance of StarGAN-VC for unseen speakers by the adoption of style embedding methods. 

In addition, Flow-based models like blow\cite{blow} are also been studied, and they transform waveforms directly rather than using acoustic features.

Both GAN and Flow-based models have in common is that they all bypass the problem of feature decoupling and convert speech directly, while there are also some other works\cite{PPG-VC1,PPG-VC2,PPG-VC-Adversarial,VQVC,VQVC+,AUTOVC} attempting to disentangle the styling unit and content unit in the embedding space. The purpose is obvious, with content information and timbre information are obtained respectively, it is easy for us to fix the content embedding while replacing the style embedding to convert the voice.

One type of methods is based on the automatic speech recognition (ASR) model \cite{PPG-VC1,PPG-VC2,PPG-VC-Adversarial,PPG3}. Firstly, a pre-trained speaker-independent ASR model was employed to extract linguistic-related features (e.g. phonetic posteriorgrams) from the source speech. Then, a synthesis model is applied to generate an utterance, of which pronunciation characteristic is very similar to the target speech. Especially in \cite{PPG-VC-Adversarial}, the pronunciation characteristic of the target speech is represented by the d-vector extracted by a pre-trained speaker recognition model, and an adversarial learning approach is used to get more pure linguistic information from phonetic posteriorgrams. 

Vector Quantization (VQ)\cite{VQVC}, a very important technology of signal compression, can quantify continuous data into discrete data. Recently, this technology has been proved that the quantized discrete data from the input continuous data is closely related with the phoneme information\cite{proveVQ}. As a result, this technology is also applied to extract the content information, and learn to obtain style embedding by the difference between original continuous space and the discrete codes. D.-Y. Wu et al. introduced this technology in the task of voice conversion. Besides, VQVC+\cite{VQVC+} proposed a new architecture combined with VQ, IN, and U-Net, which has achieved satisfactory results in the field of voice conversion.

Variational Autoencoders (VAEs)\cite{VAE1,VAE,VAE2} are also popular choices for conversion models, the network structure of VAE consists of an encoder and a decoder network. During training, the encoder learns a specific latent space from the input speech while the decoder reconstructs the speech based on that latent space. The most serious problem is that they often suffer from overly smoothed results since the fact that they impose their latent encoding to follow some prior distribution. Besides, it can also not guarantee the match between the latent distribution and the output distribution.

Recently, AutoVC\cite{AUTOVC}, a many-to-many non-parallel VC model, has gained a lot of attention due to its simple training process and simple network structure, which applies a simple vanilla autoencoder with a properly tuned information-constraining bottleneck to force disentanglement between the linguistic content and the speaker identity by training only on self-reconstruction. Compared with the previous methods, AutoVC can not only guarantee the distribution matching as GANs but also train as easily as VAEs.

Unfortunately, although the above four methods have their advantages, they all have their disadvantages. For example, in the task of feature separation of content information and voice information, those PPG-based methods have good works, but they need to introduce a complex and huge pre-training ASR network, which makes the system very complicated. On the other hand, the decoupling effect and conversion effect of VQVC is unsatisfactory. The training of AutoVC is simple enough, but its conversion effect is still not as well as expected.

In this paper, we proposed a novel voice conversion framework that combining the idea from AutoVC, text-to-speech system, and adversarial training, named TGAVC. 

Compared with AutoVC, our proposed model adds a text encoder and a speaker classifier. According to the text transcription, the text encoder produces the desired content embedding, which is used to guide the training of the content encoder. Meanwhile, the speaker classifier distinguishes speaker identity from the estimated content embedding output by the content encoder, and the adversarial training is applied between the speaker classifier and the content encoder to eliminate the speaker style information in the estimated content embedding. Experiments on AIShell-3 dataset show that our proposed model outperforms AutoVC in terms of naturalness and similarity of converted speech. The main contributions of our works as follows:

\begin{itemize}
    \item The desired content embedding produced based on the text transcription, is designed to guide the training of VC model, which makes the model more efficient and superior.
    \item The adversarial training is applied into autoencoder voice conversion model to further separate the linguistic information and the voice characteristic from speech.
    \item Compared to AutoVC, our proposed model reduced 1.72 score (the lower the better) in objective evaluation.
\end{itemize}

\section{Proposed Methods}

Our purpose is to design a many-to-many voice conversion model, which can be trained with non-parallel and multi-speaker speech datasets. We use $x_i$ to denote the speech sample, and $(t_i, u_i)$ to denote the text transcription and the corresponding speaker identity. Different from conventional VC models, our proposed method makes full use of the information of the text transcription $t_i$  to get better conversion results. For ease of presentation, the speech $x_i$ represents its acoustic features (e.g. mel-spectrogram) in this section.

\subsection{AutoVC}

\begin{figure}[t]
    \subfigure[Training]{
        \label{autovc-train}
        \begin{minipage}[b]{0.47\linewidth}
            \centering
            \includegraphics[width=0.764\linewidth]{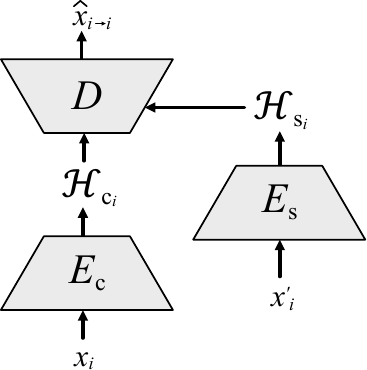}
        \end{minipage}
    }
    \subfigure[Conversion]{
        \label{autovc-conversion}
        \begin{minipage}[b]{0.47\linewidth}
            \centering
            \includegraphics[width=0.764\linewidth]{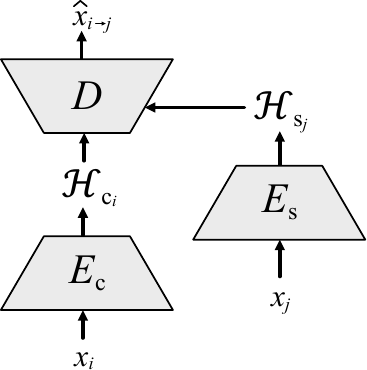}
        \end{minipage}
    }
    \caption{Framework of AutoVC.}
    \label{fig:1}
\end{figure}

AutoVC designs an autoencoder framework to solve the voice conversion task, as shown in Figure \ref{fig:1}. The framework is composed of three modules: a content encoder $E_c$ that outputs a content embedding $\mathcal{H}_c$ from source speech $x_i$, a style encoder $E_s$ that outputs a style embedding $\mathcal{H}_s$ from target speech $x_j$, and a decoder that produces the converted speech $\widehat{x}_{i \rightarrow j}$ from the content and style embeddings. In the training phase, two different speeches of the same speaker are fed into the content encoder and the style encoder respectively. Besides, the design of the self-reconstruction loss and the content code reconstruction loss enables the model to separate the content and style information from speech. 

\textbf{Self-reconstruction loss}: The self-reconstruction loss is used to render the converted speech $\widehat{x}_{i \rightarrow i}$, which is generated from the input content embedding $\mathcal{H}_c$ extracted from a source speech $x_i$ and the style embedding extracted from another speech $x'_i$ as similar as the input speech $x_i$:
\begin{equation}
    \mathcal{L}_{\text {recon }} = \mathbb{E}(\|\hat{x}_{i \rightarrow i}-x_{i}\|_{2}^{2})
\end{equation}
    
\textbf{Content code-reconstruction loss}: The aim of content code-reconstruction loss is to encourage the content encoder $E_c$ to output content embedding from input speech. What's more, it's also an effective way to guarantee that the converted speech will preserve the input composition. The content code-reconstruction loss is defined as: 
\begin{align}
    \mathcal{L}_{\text {content }} = \mathbb{E}(\|E_{c}(\hat{x}_{i
    \rightarrow i})-E_{c}(x_{i})\|_{1}) 
\end{align}

Noted that the style encoder is taken off-the-shelf, we only need to train the content encoder $E_c$ and the decoder $D$. The full objective is then formulated as:
\begin{align}
    \min _{E_{c}(\cdot), D(\cdot, \cdot)} \mathcal{L}=\mathcal{L}_{\text {recon }}+\lambda \mathcal{L}_{\text {content }}
\end{align}
\noindent where $\lambda$ denotes the weights of $\mathcal{L}_{\text {content }}$.

In conversion, we can use two speeches from different two speakers are denoted as the source and target speech. We input the source speech into the trained content encoder and the target speech into the trained style encoder, then we would get the content embedding of the source speech and the voice characteristic of the target speech, and the decoder could produce a converted speech with the linguistic information from source speech and the voice characteristic from target speech.

There is an obvious fact that the separation effect of feature decoupling is very important for all the work of speech conversion using feature decoupling. However, although AutoVC has a simple training scheme and can produce the distribution-matching voice conversion, the pure autoencoder learning process makes it difficult for us to evaluate the separation effect of content and voice characteristic.

What's more, in order to separate the content information from the input speech with only one encoder, AutoVC sets a carefully designed bottleneck dimension of the content encoder to force the encoder to disentangle the speaker-independent features. And in the decoding phase, Both content and style embedding are restored to the original temporal resolution by simple replication, we think it is not rigorous enough. Motivated by the above reasons, we focused our attention on guiding the output of the content encoder. 

\subsection{Architecture of TGAVC} 

\subsubsection{Framework}

Since the mainstream multi-speaker speech datasets provide the text transcription of each speech, we design a novel framework for voice conversion to make full use of the transcription. 

Text-to-speech(TTS) and voice conversion share the same goal to generate natural speech. Especially today, TTS models have been expanded to multi-speaker scenarios by introducing style embedding. This process is very similar to the synthesis phrase of AutoVC. In addition, since the input of the TTS models is the sequence of text, the hidden feature outputs from the encoder of TTS model must be speaker-independent. Coincidentally, the goal of AutoVC is to extract the speaker-independent feature. We naturally thought that TTS models could be used to improve the effect of AutoVC disentanglement.

\begin{figure}[t]
    \subfigure[Training]{
        \label{fig:2(a)}
        \begin{minipage}[b]{0.57\linewidth}
            \centering
            \includegraphics[width=0.9\linewidth]{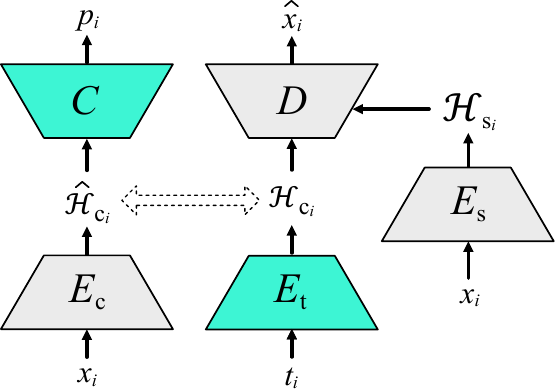}
        \end{minipage}
    }
    \subfigure[Conversion]{
        \label{fig:2(b)}
        \begin{minipage}[b]{0.37\linewidth}
            \centering
            \includegraphics[width=0.97\linewidth]{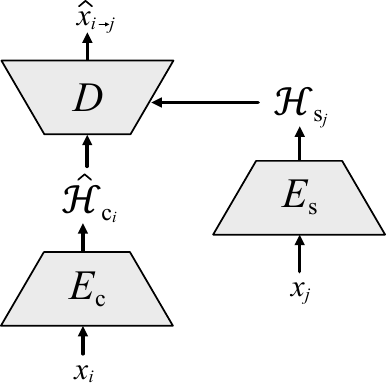}
        \end{minipage}
    }
    \caption{Framework of TGAVC.}
\end{figure}

Our proposed VC framework combines the design ideas of AutoVC, text-to-speech systems and the adversarial training. As shown in Figure~\ref{fig:2(a)}, our framework contains five modules: a content encoder $E_c$ that produces a content embedding $\widehat{\mathcal{H}}_c$ from speech, a text encoder $E_t$ that produces a text embedding $\mathcal{H}_c$ from text transcription (e.g. phonemes), we regard the output of text encoder as the ideal desired content embedding, note that there is a length regulator in the text encoder, this module is introduced to get alignment from the text embedding  $\mathcal{H}_c$ and the content embedding$\widehat{\mathcal{H}}_c$. A style encoder $E_s$ that produces a style embedding $\mathcal{H}_s$ from speech, a decoder $D$ that produces the converted speech $\widehat{x}_{i \rightarrow j}$ from the desired content and style embedding, and a speaker classifier $C$ that distinguishes speaker identity from content embedding.

In training, the speech $x_i$ is fed into the content encoder and the style encoder, the text $t_i$ is fed into the text encoder and the speaker identity $u_i$ is fed into the speaker classifier. The text encoder and the content encoder output the desired and estimated content embedding separately, which are expected to be as closed as possible. The predicted speech from the decoder is the reconstruction of the input speech $x_i$. At the same time, the speaker classifier predicts the probability of the speaker identity according to the estimated content embedding, and the adversarial training is applied between the content encoder and the speaker classifier, which can eliminate the speaker style information in the estimated content embedding. The training process is shown in Figure \ref{fig:2(a)}. 
\begin{align}
    \mathcal{H}_c &= E_t(t_i), \, \, \widehat{\mathcal{H}}_c = E_c(x_i) \\
    \hat{x}_i &= D( \mathcal{H}_c, E_s(x_i) ) \\ 
    p_i &= C(\widehat{\mathcal{H}}_c, u_i)
\end{align}

The loss function of our proposed framework consists of three-part: the reconstruction error between the predicted speech and the input speech, the estimated error between the desired and estimated content embeddings, and the adversarial loss from the speaker classifier. Instead of simply integrate all the losses into one loss function, we split them into $\mathcal{L}_1$ and $\mathcal{L}_2$. An interesting phenomenon is that in each training step, if we use $\mathcal{L}_1$ to train text encoder and decoder first, then we fix their network parameters, and then use $\mathcal{L}_2$ to update the parameters of the content encoder, the conversion effect will be much better. The loss function can be expressed as 
\begin{align}
      \mathcal{L}_1 & = \mathcal{L}_{\text {recon }}
   = \mathbb{E}(\|\hat{x}_i - x_i\|_{2}^{2}) \\   
   \begin{split}
       \mathcal{L}_2 & = \mathcal{L}_{\text {content }} + \lambda        \mathcal{L}_{\text {adv }}\\
    & = \mathbb{E}(\| \mathcal{H}_c  - \widehat{\mathcal{H}}_c \|_{1}) + \lambda \mathbb\sum_{k=1}^KI(y_{speaker}==k) \log p_k 
   \end{split}
\end{align}

\noindent where I$(\cdot)$ is the indicator function, K is the number of speakers and $y_{speaker}$ denotes speaker who produced $x_i$. Since $\lambda$ is the trade-off parameters. Then, our optimization goal is 
\begin{align}
    E_s^{*}, E_t^{*}, D = \arg \min_{E_s, E_t, D} \mathcal{L}_1
    \\
     E_c^{*} = \arg \min_{E_c} \max_{C} \mathcal{L}_2
\end{align}

In conversion, we just need to use the content encoder, the style encoder and the decoder to implement a many-to-many voice conversion system. Same as AutoVC, the source speech and the target speech are fed into the content encoder and the style encoder separately, and the decoder could produce a converted speech with the linguistic information from source speech and voice characteristic from target speech, as shown in Figure ~\ref{fig:2(b)}. 

Notice that the combination of the text encoder, the style encoder and the decoder is a multi-speaker text-to-speech system, similar to \cite{deepvoice2,fastspeech2}. In our framework, the desired content embedding from the text encoder is used to guide the training of the content encoder. Theoretically, our VC model can achieve the performance equivalent to the multi-speaker speech synthesis system if the estimated error between the desired and estimated content embedding is small enough.

\subsubsection{Network}

\begin{figure}[t]
    \centering
    \includegraphics[width=1.01\linewidth]{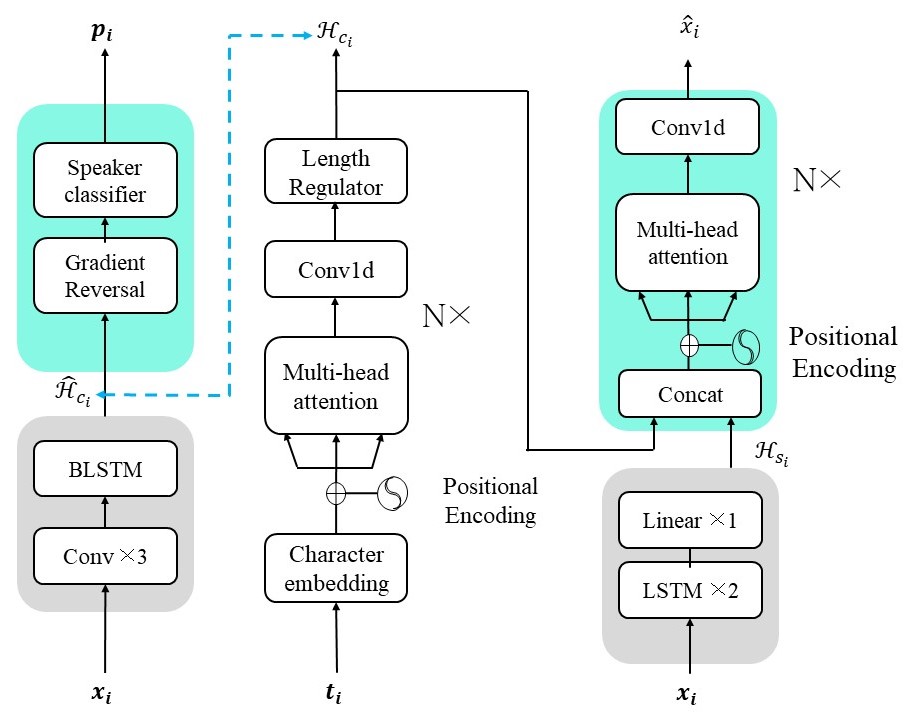}
    \caption{Network structure of TGAVC. $\widehat{\mathcal{H}}_c$  and $\mathcal{H}_c$ denotes the estimated and desired content embedding respectively. The style embedding $\mathcal{H}_s$ and $\mathcal{H}_c$ are concatenated during training. Noted that $N$ means the number of multihead-attention and Conv1d layers in encoder and decoder.  }
    \label{fig:3}
\end{figure}

The network structure of our proposed model is shown in Figure~\ref{fig:3}. The design of the text encoder and the decoder mainly draws on FastSpeech2, which is a good model based on FastSpeech\cite{fastspeech} and has shown a great effect on the TTS task. The text encoder consists of a character embedding, a stack of four FFTBlock layers which contains a multi-head attention and a two-feed-forward layer module, a full-connected layer and a Sinusoid position encoding table. this table is used to add positional embedding before we put the input feature into the FFTBlock layer stack as same as FastSpeech2. The length regulator is the same as that in FastSpeech2\cite{fastspeech2}, where the required alignment between text and speech is extracted by the Montreal forced aligner tool [MFA]. 

The style encoder is similar to that in AutoVC, which consists of a stack of two LSTM layers. Only the output of the last time is selected and fed into a fully connected layer to output the style embedding. The style encoder is pre-trained on the GE2E loss [GE2E]\cite{GE2E_loss}, which maximizes the embedding similarity among different utterances of the same speaker, and minimizes the similarity among different speaker. This loss will force the style encoder to learn features closely related to the speaker information. 

The network structure of the decoder is similar to that of the encoder,
but the difference is that the decoder has six layers of FFTBlock layers while in the encoder, the number of FFTBlock layers is four.

The style embedding is first copied to the same length as the content embedding, and then concatenated with it in the channel dimension. The concatenated embedding is passed into the decoder to generate the speech. The content encoder is composed of a stack of 3 convolution layers and a BiLSTM layer. The speaker classifier consists of a stack of two LSTM layers, a fully connected layer and a softmax layer to predict the probability for each speaker's identity.

\section{Experiments} 

\subsection{Configurations} 

In order to evaluate our proposed method, comparative experiments are conducted on AISHELL-3, a high-fidelity multi-speaker Mandarin speech corpus. This corpus contains 88035 recordings (roughly 85 hours) from 218 native Chinese mandarin speakers and includes hand-labeled full pinyin annotations. The entire dataset is randomly divided into 3 sets: 63262 recordings from 174 speakers for training, 3480 recordings for validation and other recordings for testing. It is worth noting that these voice of speakers that do not appear in training sets are used to conduct zero-shot VC experiments. In our experiments, the sampling rate of all recordings is 22.05kHz, and the mel-spectrograms are computed through a short-time Fourier transform (STFT) with Hann windowing, where 1024 for FFT size, 1024 for window size and 256 for hop size. The STFT magnitude is transformed to the mel scale using 80 channel mel filter bank spanning 90 Hz to 7.6 kHz. 

We train TGAVC model with a batch size of sixteen for 100k steps on one NVIDIA V100 GPU, and use the ADAM optimizer with $\beta_1=0.9$, $\beta_2=0.98$, $\varepsilon=10^{-9}$. The style embedding is generated by feeding 10 two-second utterances of the same speaker to the style encoder and averaging the resulting embedding. The weights in Eq.(8) are set to $ \lambda = 0.1 $. The VQVC model and AutoVC are chosen as the baseline model, of which training follows the description in \cite{AUTOVC}. In addition, for our proposed TGAVC, we propose two different training processes, the first one is that we train the content encoder jointly with the text-to-speech model. Specifically, for each training step, we first perform a text-to-speech task to train the text encoder and the decoder, and then we use the output of the text encoder of the TTS system with no grad to get the estimated error between the estimated content embedding outputs from the content encoder and the desired content embedding. The other training process is that we have trained a TTS system in advance, and then we freeze the pre-trained TTS network so that we only need to train the content encoder network in the training stage. We call this type of training "TGAVCs".

To compare the performance of different models, we used the Mel-Cepstral Distortion(MCD) between our converted speech and the ground truth target speech as our objective evaluation. Since the MCD test needs our converted speech and ground truth target speech to say the same linguistic content, we selected 10 utterances of the target speaker which have the same linguistic content as the source speaker. What's more, we also performed two subjective tests. The first test is the mean opinion score (MOS) test, where each converted utterance is rated with a score of 1-5 on the naturalness by 20 native mandarin testers. The second test is the speaker similarity test, where groups of utterances are rated with a score of 1-5 on the voice similarity. In each group, there are four converted utterances generated from AutoVC, VQVC, TGAVC and TGAVCs respectively, and one real utterance from the target speaker. Since our method does not need parallel corpus for training, the content information of ground-truth utterances from the target speaker is different from our converted speech. To evaluate the performance of Non-parallel voice conversion of our proposed method, we calculate the score according to the timbre similarity given by the tester. Thus the similarity score of 5
corresponds to the converted speech most similar to the ground-truth utterances, while the similarity score of 1 corresponds to the least similarity one. 

\subsection{Many-to-Many conversion} 

When we want to convert multiple source speakers to multiple target speakers, and both the source speaker and the target speaker are all appear in the training set, we call this type of task many-to-many conversion. We evaluate our proposed method by comparing it with AutoVC and VQVC. We did not compare it to other non-parallel many-to-many VC methods, because AutoVC has shown its advantages compared with previous work and the performance of AutoVC we used to compare is the same as that of the original work. To avoid unfair comparison, we trained AutoVC and VQVC with the same datasets. Since the style embedding of AutoVC and our proposed method are all produced by the same style encoder, it is not necessary for us to use one-hot embedding as the style embedding. 

To construct the utterances for multi many-to-many conversion evaluation, 4 speakers, 2 male and 2 female, are selected randomly from the 174 speakers in the training set. Then, we convert the test utterances of each of the 4 speakers to the other 3 speakers. So we can produce 4$\times$3 = 12 conversion utterances contains the same content information of one of the 4 speakers while carrying the other 3 speaker’s voice. Results from the MOS test and the speaker similarity test are summarized in Table~\ref{table:1}.

\vspace{-0.3cm}
\begin{table}[htbp]
   \centering
   \caption{Comparison of different methods in many-to-many VC evaluation.}
    \label{table:1}
    \begin{tabular}{l|c|c|c}
     \hline
     Method & MCD & MOS & Similarity\\
     \hline
     AutoVC &9.93±1.14&3.25±0.87&3.08±0.29 \\
     VQVC &11.04±0.82&2.64±0.67&3.15±0.68 \\
     TGAVCs &8.73±0.95&3.28±1.04&3.77±0.62 \\
     \textbf{TGAVC} &\textbf{8.21±0.97}&\textbf{3.64±0.72}&\textbf{3.84±0.49} \\
     \hline 
    \end{tabular}
\end{table}
\vspace{-0.3cm}

As is quoted in Table \ref{table:1}, the speech produced by TGAVC is much better than the baselines'. The results show that in all 4 gender groups, the MOS of TGAVC is higher than the baseline. 

In terms of similarity, TGAVC, the method we proposed, has also achieved a satisfactory result. Compared with AutoVC and VQVC, our method makes the converted speech learn better timbre information, which improves the conversion effect. The data in the table above reflect the excellent performance of TGAVC in voice conversion over the baseline. 
\subsection{Zero-shot conversion} 

In addition to good performance in traditional voice conversion tasks, the current system is also expected to have a better adaptive ability. Therefore, we carried out the zero-shot conversion task. That is, the target speaker is unseen in the training set, and only a few utterances of each target speaker are available in inference.

In the past, many methods usually used one-hot coding as style embedding, which makes them unable to encode unseen speakers to do zero-shot conversion, but AutoVC, VQVC and our proposed TGAVC are special, by using a learnable style embedding, we can easily accomplish the conversion task. The baseline we used to compare are still AutoVC and VQVC.

For one-shot conversion evaluation, we select a few speakers from the test set who have never appeared in the training set as our target timbre. To get the target style embedding, we input 10 utterances of the target speaker into the trained style encoder. Since it is difficult for us to find parallel data in the dataset, we no longer conduct objective analysis. Besides, for a more comprehensive comparison of our proposed method and AutoVC conversion effect, we divide the converted audio into two gender groups, the one is same-sex transformation, including male to male and female to female, the other one is opposite-sex transformation, including female to male and male to female, and we summarized our results of conversion evaluation in Table~\ref{table:2}.

\vspace{-0.3cm}
\begin{table}[htbp]
    \centering
   \caption{Comparison of different methods in zero-shot VC evaluation. (SIM:Similarity)}
    \label{table:2}
    \begin{tabular}{l|c|c|c}
     \hline
     Method & MOS & same-sex SIM & oppo-sex SIM \\
     \hline
     AutoVC & 3.04±1.07 & 2.88±0.29 & 3.01±0.45 \\
     VQVC & 2.49±0.73 & 3.15±0.56 & 3.52±0.39 \\
     TGAVCs  & 3.17±0.96 & 3.42±0.67 & 3.33±0.67 \\
     \textbf{TGAVC} &\textbf{3.42±1.13} &\textbf{3.64±0.52} &\textbf{3.62±0.39} \\
     
     \hline 
    \end{tabular}
\end{table}
\vspace{-0.3cm}

The result shows that even for unseen speakers, the proposed method is still better than the baseline in natural evaluation. In addition, compared with baseline synthesized speech, there are lots of people who think that the synthesized speech conversion by our method is more similar to the ground truth, which demonstrates TGAVC’s competence in zero-shot conversion. To compare the differences between our method and baseline more intuitively, we draw the  MOS Score results into the following histogram. 

\begin{figure}
    \centering
    \begin{tikzpicture}
    \tikzstyle{every node}=[font=\fontsize{7.5}{3}\selectfont]
\begin{axis}
[
    ybar,
    x=58pt,
    y=60pt,
    enlargelimits=0.15, 
    legend style={at={(0.58, 1.15)}, 
      anchor=north, legend columns=100}, 
    symbolic x coords={F-F, F-M, M-F, M-M}, xtick=data, nodes near coords, nodes near coords align={vertical}, ]
\addplot[fill=green] coordinates {(F-F, 3.45) (F-M, 3.075) (M-F, 2.95) (M-M, 2.7)}; 
\addplot coordinates {(F-F, 2.85) (F-M, 2.475) (M-F, 2.45) (M-M, 2.2)};
\addplot[fill=cyan] coordinates {(F-F, 3.875) (F-M, 3.3) (M-F, 3.75) (M-M, 2.75)};
\addplot[fill = orange] coordinates {(F-F, 3.71) (F-M, 2.95) (M-F, 3.52) (M-M, 2.49)};

\legend{AutoVC, VQVC, TGAVC, TGAVCs}

\end{axis}
\end{tikzpicture}
    \caption{MOS score results for zero-shot conversion.}
    \label{fig:4}
\end{figure}
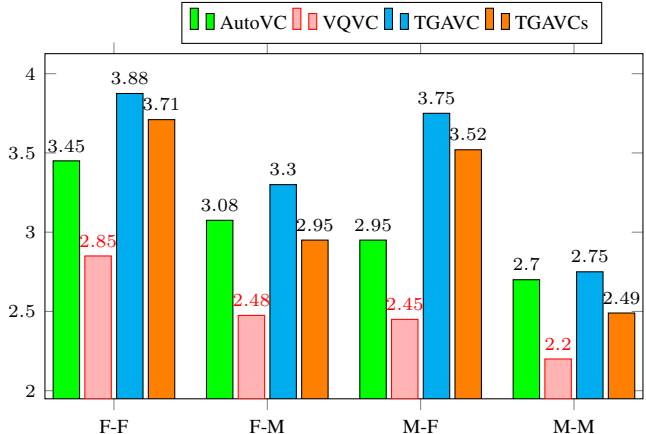

\subsection{Analysis} 

In this section, we will focus on the performance differences between TGAVC and TGAVCs. As is shown in the Figure~\ref{fig:4}, both TGAVC and TGAVCs have better performance when they are converted to female voice, but worse when the target speaker is male, which may be caused by the imbalance of samples. By comparison, The proposed TGAVC is always better than TGAVCs while the training time of TGAVC is less than that of the latter. That is why we choose TGAVC.  

It is worth noting that we have considered using only one optimizer to optimize all network modules at the same time, but as we mentioned above, this kind of training method makes the result of VC synthesizes very poor. We estimate that this may be because when we update the parameters of the content encoder and text encoder at the same time to make their outputs as close as possible, the content encoder will finally learn an intermediate vector between desired content embedding and estimated content embedding instead of the expected content embedding. 

\section{Conclusion}

We have proposed TGAVC, a novel VC system combining AutoVC and the TTS system. During training, the content encoder output is guided to become more and more similar to the text embedding outputs from the encoder of the TTS system. We also introduce an adversarial learning approach to improve speaker identity in non-parallel many to many voice conversion. The encoder output is optimized to be more speaker-independent to improve the decoupling of content and timbre of input speech. We conducted both objective and subjective evaluation on traditional many to many VC and one-shot VC. The result shows that the method we proposed significantly improves the acoustic quality of the synthesized speech in VC task and the similarity with the target voice under both conditions.

\bibliographystyle{IEEEbib}

\bibliography{refs}

\end{document}